\documentclass[dvips]{acta}
\usepackage{supertabular,lscape,epsfig}
\usepackage{amssymb}
\usepackage{amsmath}

\newcommand\farcs{\hbox{$.\!\!^{\prime\prime}$}}
\newcommand\fs{\hbox{$.\!\!^{\scriptstyle\mathrm{s}}$}}

\newcommand{\figLCs}{1}
\newcommand{\figCVdistr}{2}
\newcommand{\figCVampl}{3}
\newcommand{\figCVstats}{4}
\newcommand{\figWZSge}{5}
\newcommand{\figSUUMa}{6}
\newcommand{\figZCam}{7}
\newcommand{\figBLG}{8}
\newcommand{\tabOGLEDN}{1}
\newcommand{\tabWZSge}{2}
\newcommand{\tabZCam}{3}

\SetPages{0}{0}

\SetVol{58}{2008}

\begin{document}

\begin{Titlepage}
\Title{One Thousand New Dwarf Novae from the OGLE Survey}

\Author{P.~Mr{\'o}z$^1$, A.~Udalski$^1$, R.~Poleski$^{1,2}$, P.~Pietrukowicz$^{1}$,\\ M.~K.~Szyma\'nski$^{1}$, I.~Soszy\'nski$^{1}$, \L{}.~Wyrzykowski$^{1}$, K.~Ulaczyk$^{1,3}$, S.~Koz\l{}owski$^{1}$ and J.~Skowron$^{1}$}
{$^1$Warsaw University Observatory, Al. Ujazdowskie 4, 00-478 Warsaw, Poland\\
e-mail: pmroz@astrouw.edu.pl\\
$^2$Department of Astronomy, Ohio State University, 140 W. 18th Ave., Columbus, OH 43210, USA \\
$^3$Department of Physics, University of Warwick, Coventry CV4 7AL, UK
}

\Received{Month Day, Year}
\end{Titlepage}

\Abstract{

We present one of the largest collections of dwarf novae (DNe) containing 1091 objects that have been discovered in the long-term photometric data from the Optical Gravitational Lensing Experiment (OGLE) survey. They were found in the OGLE fields toward the Galactic bulge and the Magellanic Clouds. We analyze basic photometric properties of all systems and tentatively find a population of DNe from the Galactic bulge. We identify several dozen of WZ Sge-type DN candidates, including two with superhump periods longer than 0.09 d. Other interesting objects include SU UMa-type stars with ``early'' precursor outbursts or a Z Cam-type star showing outbursts during standstills. We also provide a list of DNe which will be observed during the K2 Campaign 9 microlensing experiment in 2016. Finally, we present the new OGLE-IV real-time data analysis system: CVOM, which has been designed to provide continuous real time photometric monitoring of selected CVs. 
}{
novae, cataclysmic variables
}

\section{Introduction}

Dwarf nova (DN) outbursts are one of the most common groups of Galactic transients. They are caused by instabilities in accretion disks in cataclysmic binary systems (CVs, \eg Warner 1995). During outbursts that last from a few to several days, DNe can increase their brightness by 8 mag. Outbursts recur on time scales from days to several years. Properties of DN outbursts (frequency, amplitude, duration, etc.) depend on physical parameters of the underlying binary, such as orbital period, mass-transfer rate, or magnetic field of the white dwarf. Thus, if one wants to unravel statistical properties of CV population, a large sample of objects with well-defined selection criteria is necessary.

For many years, however, such a sample was not present. Most of the known objects were relatively close (and bright) or had large outburst amplitudes. A large number of transients, including DNe, is recently announced by large-scale sky surveys, such as the Catalina Real-time Transient Survey (CRTS, Drake \etal 2009, 2014), the Panoramic Survey Telescope and Rapid Response System (PanSTARRS, Hodapp \etal 2004), the Intermediate Palomar Transient Factory (iPTF, Kulkarni 2013), or the All-Sky Automated Survey for Supernovae (ASAS-SN, Shappee \etal 2014). However, because of technical and seeing limitations those surveys cannot observe the densest regions of the Galactic disk and bulge, where the majority of observable CVs is believed to be located. 

The Optical Gravitational Lensing Experiment (OGLE, Udalski \etal 2015) is one of the largest sky variability surveys. During its fourth phase (OGLE-IV, since 2010), an area of over 3500 square degrees (the Galactic bulge and disk, the Magellanic System) is regularly monitored. Up to now, dozens of CVs were reported from the OGLE-II data (Cieslinski \etal 2003) and the OGLE-III Galactic disk fields (Mr\'oz \etal 2013). Poleski \etal (2011) presented the study of three systems located in the high-cadence OGLE-IV fields. 
In this paper we present one of the largest uniform samples of DN candidates which were detected in the OGLE photometric data up to 2013 December 31st. The time span of analyzed OGLE observations is up to twenty years and the cadence as short as twenty minutes. We analyze the basic properties of these systems (outburst frequency, amplitude, duration, orbital and superhump periods etc.), and present several intriguing objects. We also provide a list of CVs which will be observed during the K2 Campaign 9 microlensing experiment in 2016.

\section{Data}
Photometric data that were analyzed in this paper were collected in the course of two phases of the OGLE survey: OGLE-III (2001-2009, Udalski \etal 2008) and OGLE-IV (since 2010, Udalski \etal 2015). Additionally, light curves of several dozens of objects are supplemented by photometry from the OGLE-I (1992-1995, Udalski \etal 1992) and OGLE-II (1997-2000, Udalski \etal 1997). For thirty objects our photometry spans over twenty years. The cadence of observations varies from 20 min in the central Galactic bulge regions to 2--5 d in the Magellanic Clouds. The majority of measurements were taken through the $I$-band filter, closely resembling that of the standard Cousins system. The photometry was carried out with the Difference Image Analysis (DIA) algorithm (Alard and Lupton 1998, Wo\'zniak 2000). The photometry of faint objects was improved by re-determining their centroids on the subtracted images and then re-reducing the data with the standard pipeline. For some faint objects in crowded fields, there might a small (up to 0.1--0.2 mag) systematic shift between photometry from different phases of the project, which is likely caused by different blending on the reference images. The photometry reaches $I \approx 20.5-21.7$ mag, depending on the stellar crowding of the field. 

\subsection{Sample Selection}
The final sample consists of 1059 objects in the Galactic bulge fields and 32 stars located toward the Magellanic System (the latter stars belong to the Milky Way, DNe from the Magellanic Clouds would be much fainter). They were selected using two methods. 789 stars were detected as DN candidates by the Early Warning System (EWS, Udalski 2003) which has been used for the detection of microlensing events. DN outbursts are easily distinguishable from microlensing events because they brighten faster and sharper than the fading part. However, DNe with frequent outbursts cannot be detected by the EWS. Hence, we additionally analyzed light curves of all objects from the OGLE Galactic bulge and Magellanic Clouds' fields with an algorithm presented in Mr\'oz \etal (2013). In short, we searched for brightenings by at least 1.0 mag for at least three consecutive nights and the selected candidates were visually inspected. We found 679 DNe (377 were identified both by the above algorithm and the EWS). In Table \tabOGLEDN{}, we present basic parameters of all DN candidates (the full table is available online). Exemplary light curves are shown in Fig. \figLCs{}.

We cross-matched our list with the International Variable Star Index (VSX) database (Watson 2006): 98 objects were previously known, most ($\approx 80\%$) of which were discovered in data from the MACHO (Cieslinski \etal 2004) and OGLE (Cieslinski \etal 2003, Poleski \etal 2011, Udalski \etal 2012) surveys. A few other systems were listed in Downes \etal (2001) catalog. The complete list of cross-identifications and references is available online.

\subsection{Online Data}

The final long-term $I$-band light curves of all of the objects presented in this paper are available to the astronomical community from the OGLE Internet Archive:
\begin{center}
{\it
http://ogle.astrouw.edu.pl \\
ftp://ftp.astrouw.edu.pl/ogle/ogle4/OCVS/CV/
}
\end{center}

The objects are arranged according to increasing right ascension\footnote{With the exception of OGLE-BLG-DN-0001, OGLE-BLG-DN-0002, and OGLE-BLG-DN-0003, which were previously reported by Poleski \etal (2011).} and named as OGLE-BLG-DN-NNNN (stars in the Galactic bulge fields) or OGLE-MC-DN-NNNN (stars toward the Magellanic Clouds), where NNNN is a consecutive number. 

The real-time photometry of selected objects brighter than $I_{\rm max} < 17$ mag will be available via the OGLE Cataclysmic Variable Monitoring System (see Section~6).

\begin{landscape}
\renewcommand{\TableFont}{\scriptsize}
\MakeTable{@{}rr@{\uph}r@{\upm}r@{\fs}lr@{\arcd}r@{\arcm}r@{\farcs}lrrrrrrrrrr@{}}{\textwidth}{OGLE DN candidates (first 25 objects).}
{
\hline
\multicolumn{1}{c}{Name} & \multicolumn{4}{c}{RA} & \multicolumn{4}{c}{Decl.} & \multicolumn{1}{c}{Field} & \multicolumn{1}{c}{Star ID} & \multicolumn{1}{c}{Field} & \multicolumn{1}{c}{Star ID} & \multicolumn{1}{c}{Field} & \multicolumn{1}{c}{Star ID} & \multicolumn{1}{c}{$I_{\rm max}$}  & \multicolumn{1}{c}{$\Delta I$} & \multicolumn{1}{c}{$\nu_{\rm obs}$} & \multicolumn{1}{c}{$\langle D\rangle$} \\
 & \multicolumn{4}{c}{J2000.0} & \multicolumn{4}{c}{J2000.0} & \multicolumn{1}{c}{(OGLE-IV)} & \multicolumn{1}{c}{(OGLE-IV)} & \multicolumn{1}{c}{(OGLE-III)} & \multicolumn{1}{c}{(OGLE-III)} & \multicolumn{1}{c}{(OGLE-II)} & \multicolumn{1}{c}{(OGLE-II)} & \multicolumn{1}{c}{[mag]} & \multicolumn{1}{c}{[mag]} & \multicolumn{1}{c}{[yr$^{-1}$]} & \multicolumn{1}{c}{[d]}\\
\hline
OGLE-BLG-DN-0001 & 17 & 53 & 10 & 04 & -29 & 21 & 20 & 4 & BLG501.27 & 190803 & BLG195.7 & 36348  & - & -              & 15.71 & 2.65 & 3.89 & 10.3 \\
OGLE-BLG-DN-0002 & 17 & 53 & 18 & 72 & -29 & 17 & 17 & 5 & BLG501.27 & 145001 & BLG195.7 & 145559 & - & -              & 16.18 & 2.53 & 0.49 & 14.4 \\
OGLE-BLG-DN-0003 & 17 & 56 & 07 & 29 & -27 & 48 & 47 & 4 & BLG504.23 & 109722 & -        & -      & - & -              & 16.06 & 2.17 & 0.38 & 16.3 \\
OGLE-BLG-DN-0004 & 17 & 13 & 59 & 69 & -29 & 20 & 00 & 0 & BLG617.32 & 35044  & BLG363.4 & 84924  & - & -              & 18.69 & 1.54 & 3.15 & 9.2 \\
OGLE-BLG-DN-0005 & 17 & 14 & 11 & 52 & -29 & 29 & 43 & 0 & BLG617.24 & 23320  & BLG363.3 & 90335  & - & -              & 17.73 & 3.02 & 7.32 & 6.9 \\
OGLE-BLG-DN-0006 & 17 & 14 & 37 & 09 & -29 & 37 & 19 & 4 & BLG617.23 & 23N    & -        & -      & - & -              & 18.02 & 2.17 & 3.31 & 9.1 \\
OGLE-BLG-DN-0007 & 17 & 14 & 42 & 64 & -29 & 43 & 45 & 1 & BLG617.14 & 58505  & BLG363.2 & 23794  & - & -              & 14.34 & 2.40 & 0.48 & 50.0 \\
OGLE-BLG-DN-0008 & 17 & 15 & 48 & 63 & -29 & 31 & 10 & 9 & BLG617.21 & 114347 & -        & -      & - & -              & 17.19 & 2.32 & 6.61 & 5.4 \\
OGLE-BLG-DN-0009 & 17 & 16 & 02 & 98 & -29 & 10 & 10 & 1 & BLG617.29 & 55358  & BLG359.8 & 118520 & - & -              & 16.47 & 1.67 & 0.73 & 16.0 \\
OGLE-BLG-DN-0010 & 17 & 17 & 17 & 93 & -29 & 50 & 25 & 2 & BLG617.10 & 75166  & BLG362.1 & 56666  & - & -              & 18.07 & 1.96 & 1.84 & 12.6 \\
OGLE-BLG-DN-0011 & 17 & 17 & 26 & 03 & -28 & 33 & 25 & 1 & BLG616.10 & 55156  & -        & -      & - & -              & 17.75 & 2.53 & 7.61 & 8.6 \\
OGLE-BLG-DN-0012 & 17 & 21 & 19 & 07 & -30 & 04 & 25 & 0 & BLG615.16 & 86358  & -        &  -     & - & -              & 17.46 & 2.43 & 11.88 & 7.2 \\
OGLE-BLG-DN-0013 & 17 & 22 & 01 & 55 & -29 & 18 & 43 & 0 & BLG615.32 & 93733  & -        & -      & - & -              & 17.14 & 3.04 & 12.62 & 6.4 \\
OGLE-BLG-DN-0014 & 17 & 24 & 59 & 44 & -30 & 06 & 39 & 3 & BLG615.11 & 79005  & -        & -      & - & -              & 18.66 & 1.89 & 3.73 & 12.7 \\
OGLE-BLG-DN-0015 & 17 & 25 & 00 & 02 & -29 & 21 & 29 & 2 & BLG615.28 & 96289  & BLG331.2 & 156662 & - & -              & 18.73 & 0.76 & 3.01 & 9.7 \\
OGLE-BLG-DN-0016 & 17 & 25 & 52 & 71 & -29 & 56 & 12 & 7 & BLG615.10 & 49978  & -        & -      & - & -              & 15.84 & 2.82 & 22.76 & 7.6 \\
OGLE-BLG-DN-0017 & 17 & 26 & 21 & 17 & -28 & 15 & 58 & 0 & BLG614.26 & 58992  & -        & -      & - & -              & 14.72 & 2.40 & 10.18 & 6.1 \\
OGLE-BLG-DN-0018 & 17 & 26 & 28 & 16 & -29 & 24 & 49 & 3 & BLG615.26 & 74186  & -        & -      & - & -              & 18.20 & 0.81 & 2.60 & 9.0 \\
OGLE-BLG-DN-0019 & 17 & 26 & 53 & 38 & -30 & 15 & 16 & 4 & BLG615.01 & 15248  & -        & -      & - & -              & 15.59 & 3.67 & 0.60 & 14.1 \\
OGLE-BLG-DN-0020 & 17 & 26 & 59 & 81 & -29 & 07 & 58 & 5 & BLG613.25 & 99667  & -        & -      & - & -              & 15.57 & 0.58 & 0.37 & 7.8 \\
OGLE-BLG-DN-0021 & 17 & 27 & 02 & 77 & -40 & 09 & 12 & 1 & -         & -      & BLG200.1 & 26109  & BUL\_SC47 & 334377 & 17.43 & 4.48 & 1.73 & 18.7 \\
OGLE-BLG-DN-0022 & 17 & 27 & 52 & 65 & -29 & 48 & 38 & 3 & BLG613.07 & 38625  & -        & -      & - & -              & 17.37 & 2.51 & 0.37 & 19.9 \\
OGLE-BLG-DN-0023 & 17 & 28 & 04 & 58 & -27 & 43 & 04 & 3 & BLG612.24 & 68N    & -        & -      & - & -              & 18.25 & 1.94 & 7.11 & 6.8 \\
OGLE-BLG-DN-0024 & 17 & 28 & 50 & 69 & -27 & 24 & 14 & 4 & BLG612.31 & 13827  & -        & -      & - & -              & 18.75 & 1.53 & 1.12 & 15.4 \\
OGLE-BLG-DN-0025 & 17 & 29 & 03 & 05 & -29 & 30 & 57 & 3 & BLG613.13 & 409N   & -        & -      & - & -              & 17.81 & 2.39 & 0.37 & 24.6 \\
\hline
\multicolumn{19}{p{19cm}}{The full table is available online. For each object, we list designation (Col. 1), equatorial coordinates (for the epoch J2000.0, Cols. 2-3), OGLE database identifiers (Cols. 4-9), peak magnitude $I_{\rm max}$ (Col. 10), amplitude $\Delta I$ of outbursts (Col. 11), outburst frequency $\nu_{\rm obs}$ (Col. 12), and mean outburst duration $\langle D\rangle$ (Col. 13).}
}
\end{landscape}

\begin{figure}[htb]
\includegraphics[width=\textwidth]{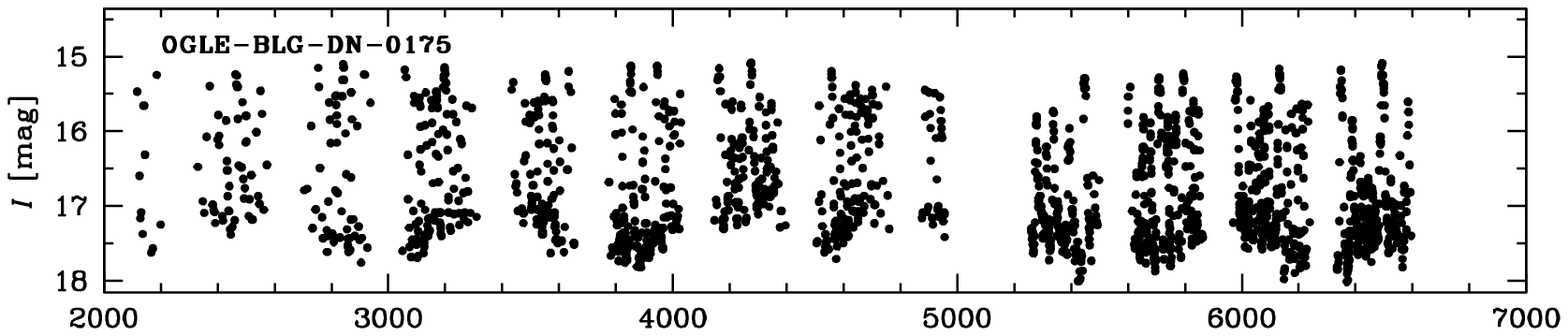}\\
\includegraphics[width=\textwidth]{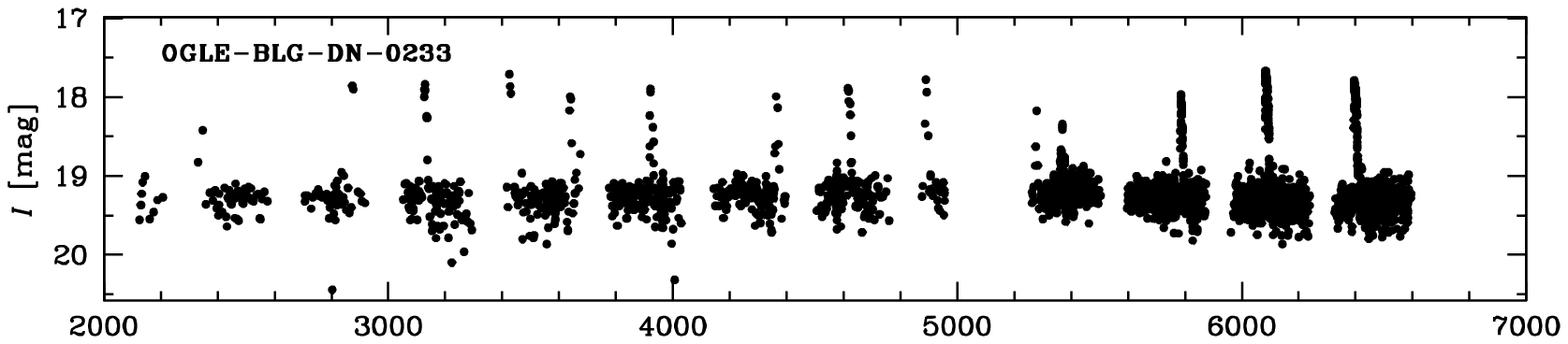}\\
\includegraphics[width=\textwidth]{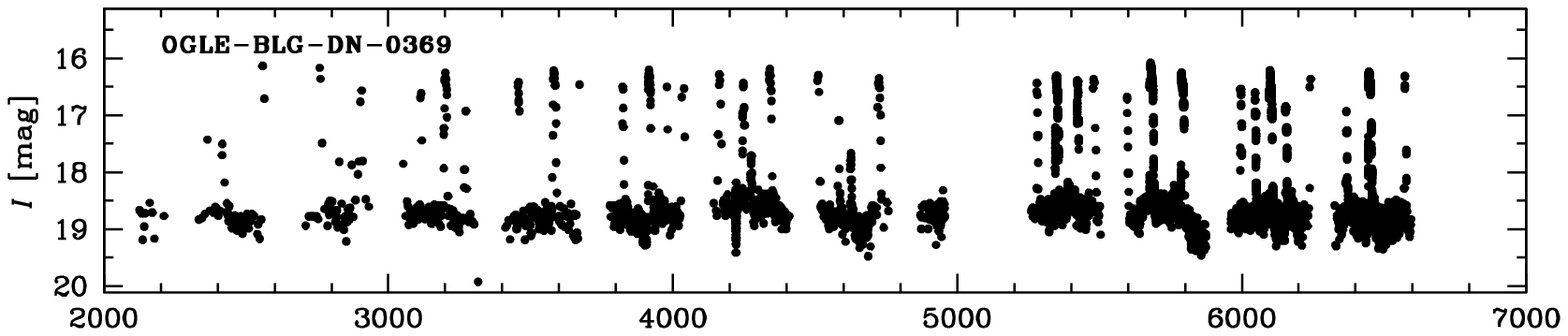}\\
\includegraphics[width=\textwidth]{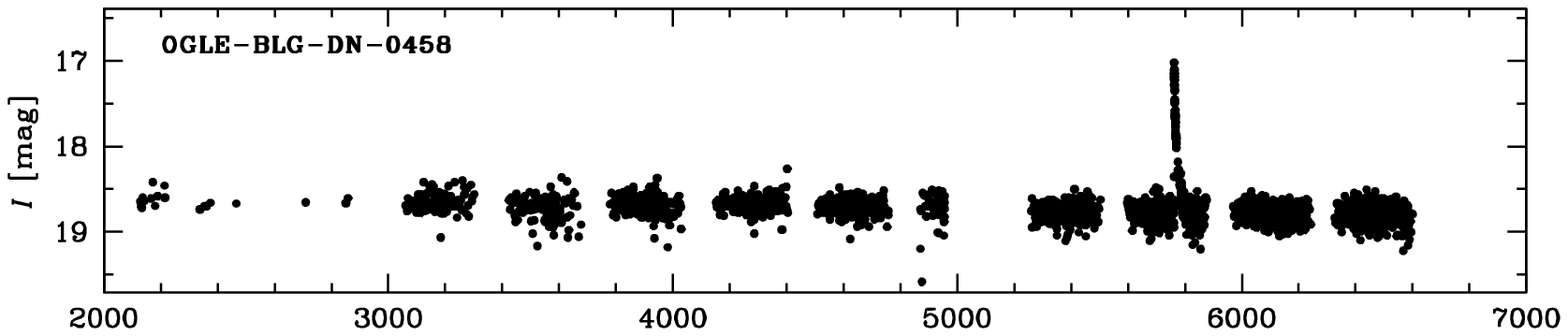}\\
\includegraphics[width=\textwidth]{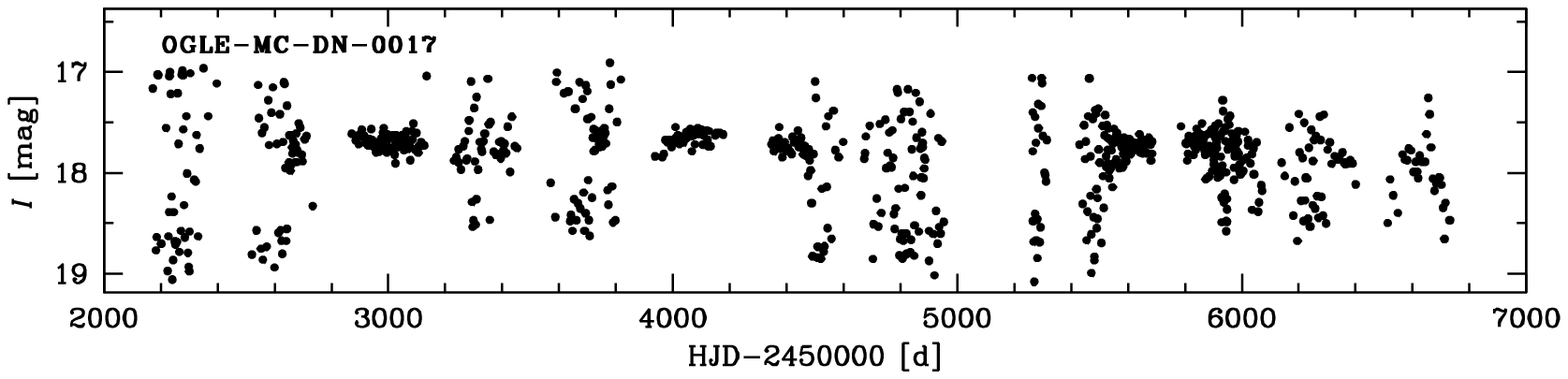}\\
\FigCap{Examples of long-term light curves of dwarf novae.}
\end{figure}

\section{Results}

First, we have to address the question of purity of our sample. Because outbursts are generally well-covered (from a few to several dozen data points per outburst) and DNe have characteristic light curves, the chance of misclassification is small. However, our classification might be ambiguous for faint, low-amplitude objects that have sparse temporal sampling, but their number does not exceed 5--10 per cent of the total sample. There are several transient classes that may contaminate this collection:
\begin{itemize}
\item extragalactic transients (supernovae, AGNs, etc.): they are generally not expected at low Galactic latitudes, because of crowding and high interstellar reddening, although we have seen a few bright supernovae ($I\approx 17$ mag at the peak) at $|b|\lesssim 5^{\circ}$;
\item classical novae: bright and easy to distinguish, although some distant and/or heavily reddened (fainter than $\approx 18$ mag at the peak) novae might have been mistaken with DNe;
\item X-ray binaries: amplitudes of optical outbursts can reach up to a few magnitudes (\eg Chen \etal 1997);
\item young stars: which show a wide range of light curve amplitudes and morphologies, including quasi-periodic outbursts (\eg Cody \etal 2014);
\item Be stars: they exhibit low-amplitude variability (up to $\sim 1.2$ mag in extreme cases, \eg Keller \etal 2002); 
\item microlensing events: single microlenses have characteristic symmetric light curves, which are easy to recognize; binary lensing events with the second caustic crossing not covered may be mistaken with DNe.
\end{itemize}

Figure \figCVdistr{} shows the spatial distribution of DN candidates (with outburst amplitudes of at least one mag). As expected, the number of discovered objects depends strongly on cadence and time span of observations. Fields in the southern Galactic hemisphere were monitored with high cadence (up to 30 times per night), while those along the Galactic equator were observed occasionally. Thus, for possible DN outbursts we have only single data points, which prevents robust detection and characterization.

Figure \figCVampl{} shows the distribution of peak magnitudes and outburst amplitudes. Similarly as for DNe in the OGLE-III Galactic disk fields (Mr\'oz \etal 2013), outburst amplitudes seem to be somewhat lower than measured in other surveys (for example the histogram of outburst amplitudes for CRTS CVs is symmetric with the mean of $\approx 3$ mag and the FWHM of $\approx 3$ mag; see Fig. 9 of Drake \etal 2014 for comparison). First, because CVs are intrinsically faint in quiescence, we might not observe ``true'' quiescence because of blending. This is supported by measured amplitudes (average amplitude of 3.58 mag with dispersion of 1.04 mag) of DNe toward the Magellanic Clouds, where the density of stars is much lower and the blending becomes unimportant. Moreover, because the spectral energy distribution of DNe in outburst resemble that of early-type stars and our observations are taken in the $I$-band, amplitudes will be slightly smaller than in visual bands.  Finally, some of ``stunted'' outbursts might be real (\eg Honeycutt \etal 1998, Schreiber \etal 2000).

\begin{figure}[htb]
\includegraphics[width=0.98\textwidth]{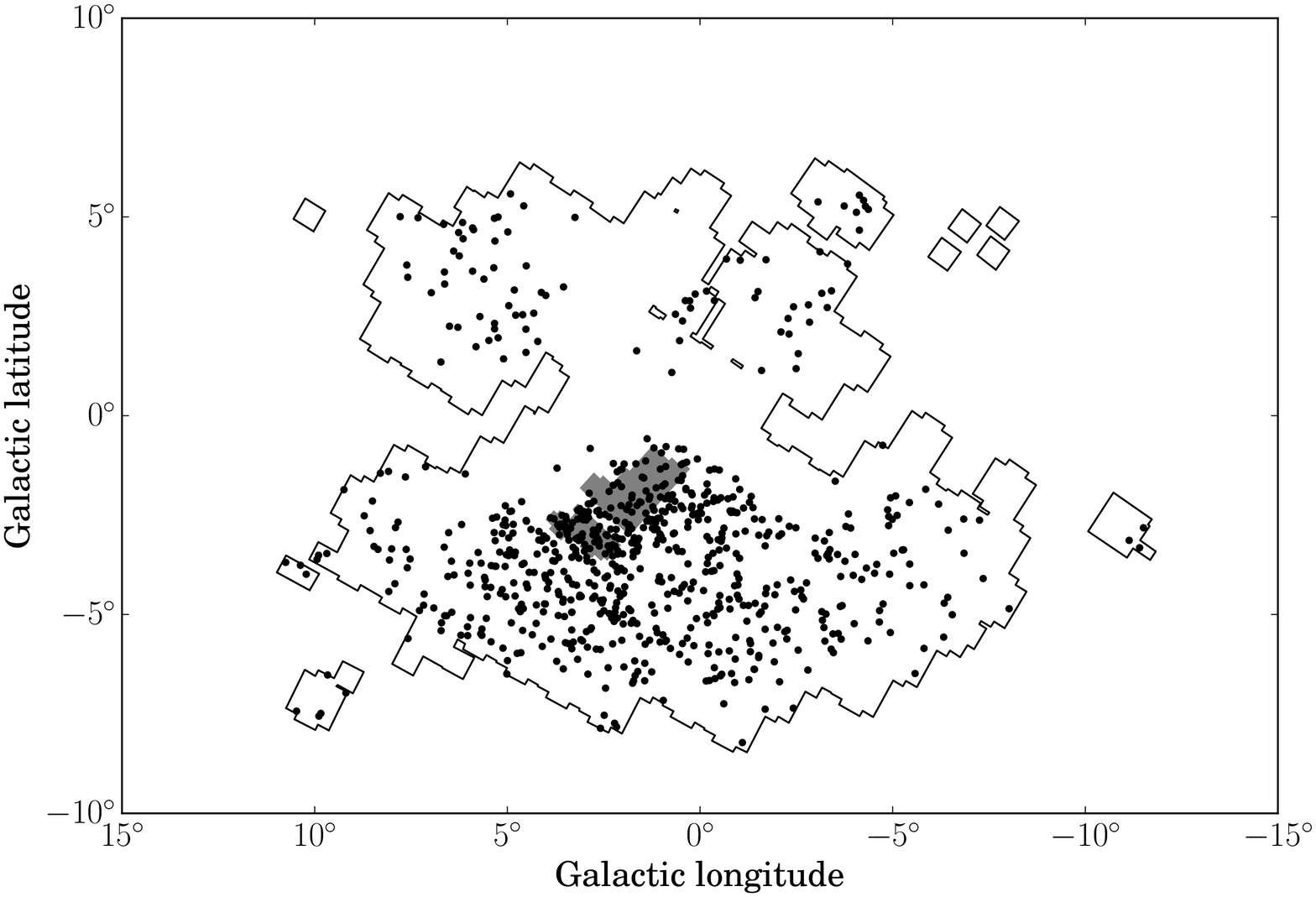}
\FigCap{Distribution of DN candidates with outbursts amplitudes larger than 1 mag in Galactic coordinates. Black lines show contours of the OGLE fields toward the Galactic bulge. Fields along the Galactic equator were observed at low cadence which prevents robust detection of DN outbursts. Shaded area is the K2 Campaign 9 superstamp (Section 5).}
\end{figure}

\begin{figure}[htb]
\includegraphics[width=0.49\textwidth]{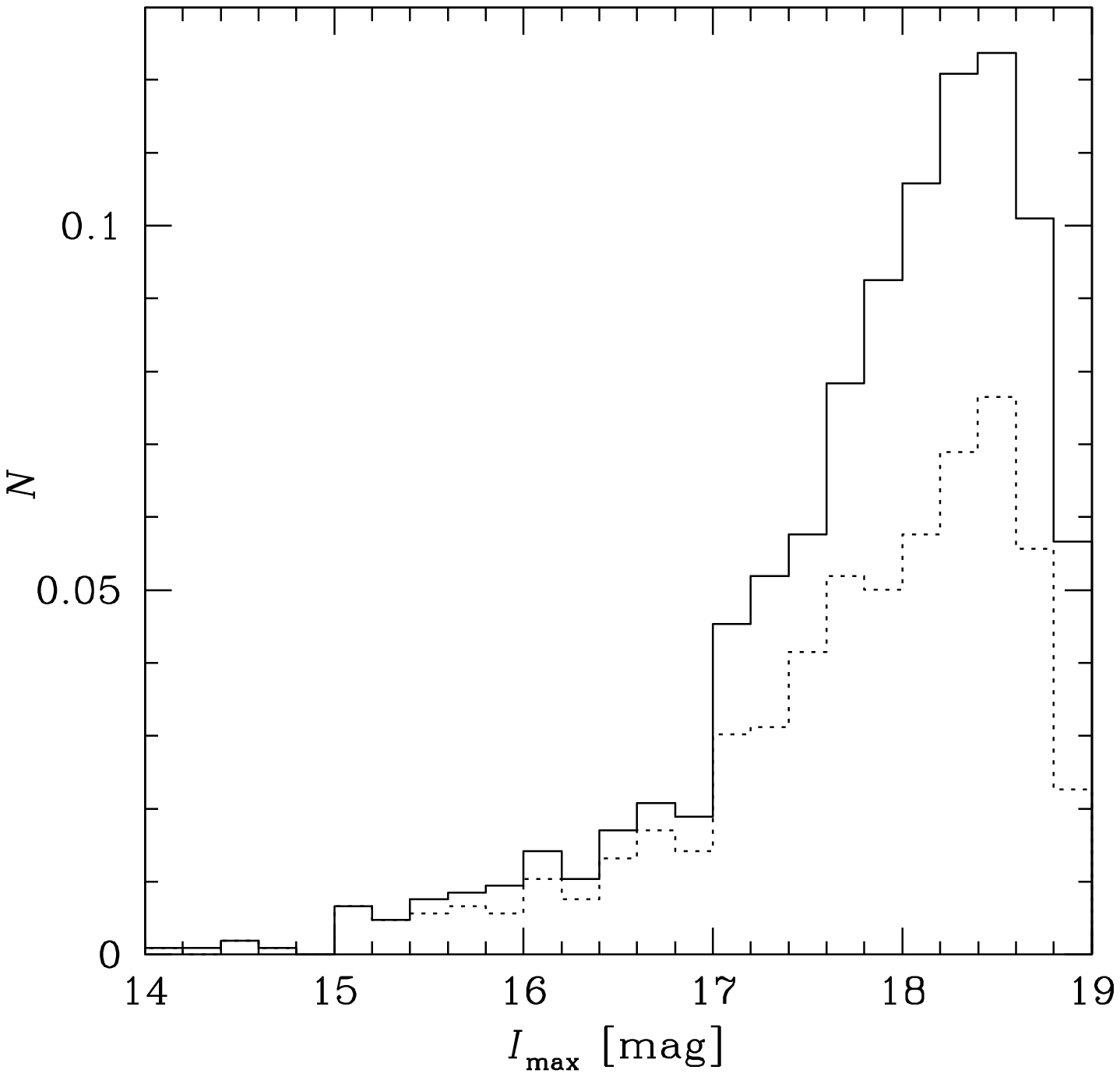}
\includegraphics[width=0.49\textwidth]{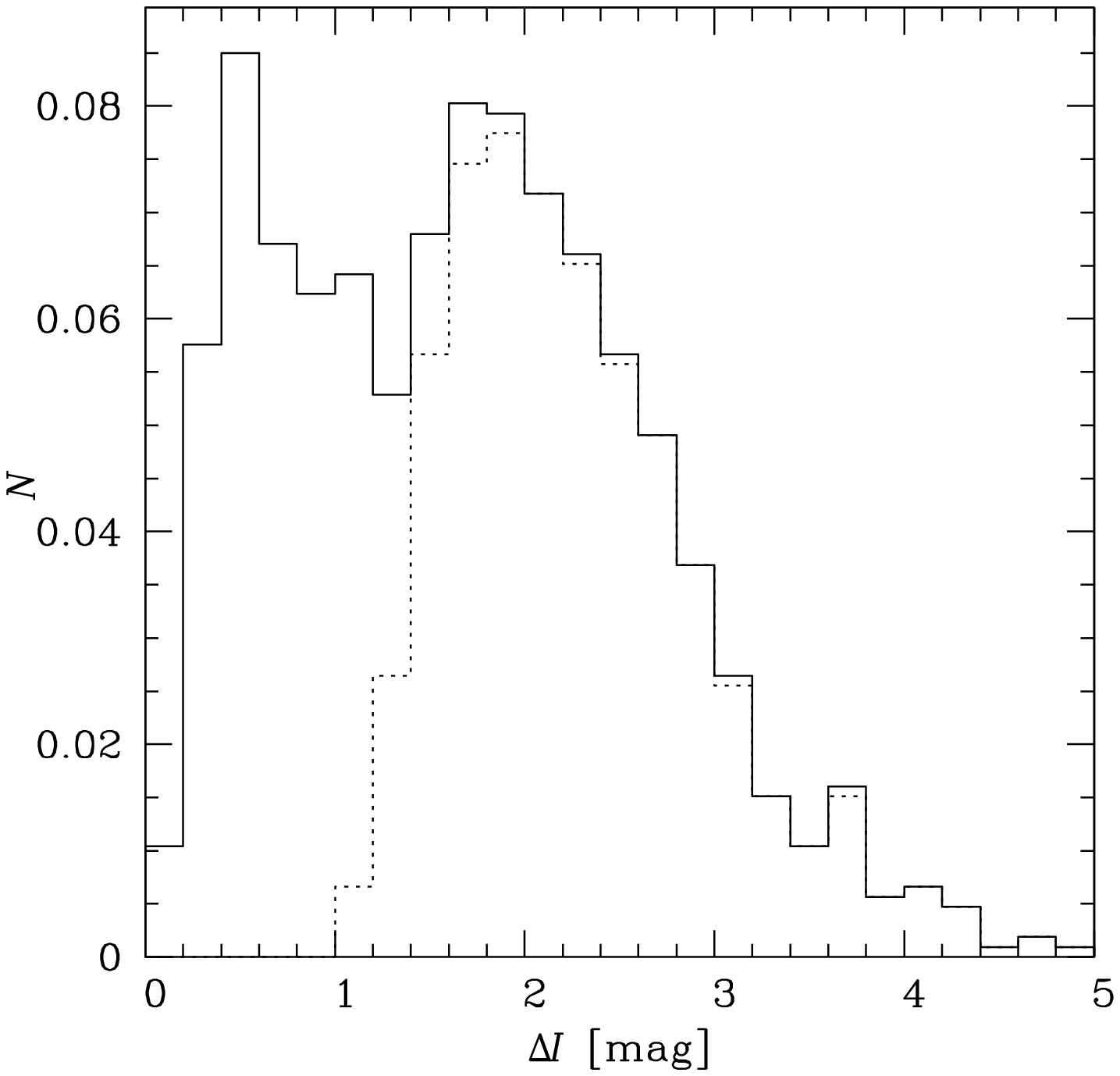}
\FigCap{Distribution of peak magnitudes ({\it left}) and outburst amplitudes ({\it right}) of dwarf novae toward the Galactic bulge. DNe discovered by the algorithm of Mr\'oz \etal (2013) are given by the dotted line. Note that the number of faint, low-amplitude DNe increases significantly when $I_{\rm max} > 17$ mag -- these may be Galactic bulge objects.}
\end{figure}

We point out that the number of stars with low-amplitude outbursts increases significantly when $I_{\rm max} > 17$ mag, as shown in the left panel of Figure~\figCVampl{}. That excess may be produced by DNe located in the Galactic bulge, for which only outburst peaks may be observed.

\begin{figure}[htb]
\includegraphics[width=0.49\textwidth]{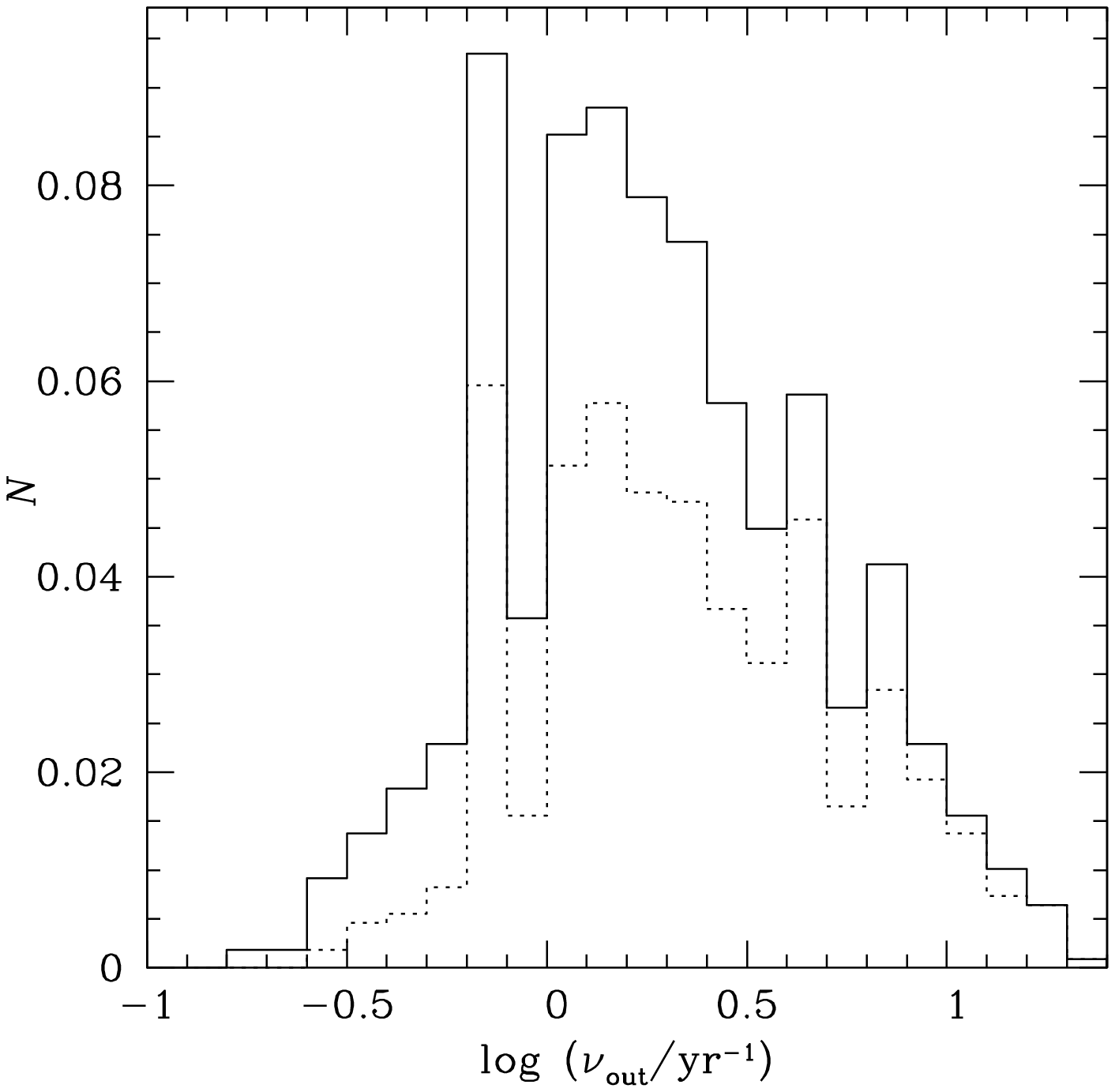}
\includegraphics[width=0.49\textwidth]{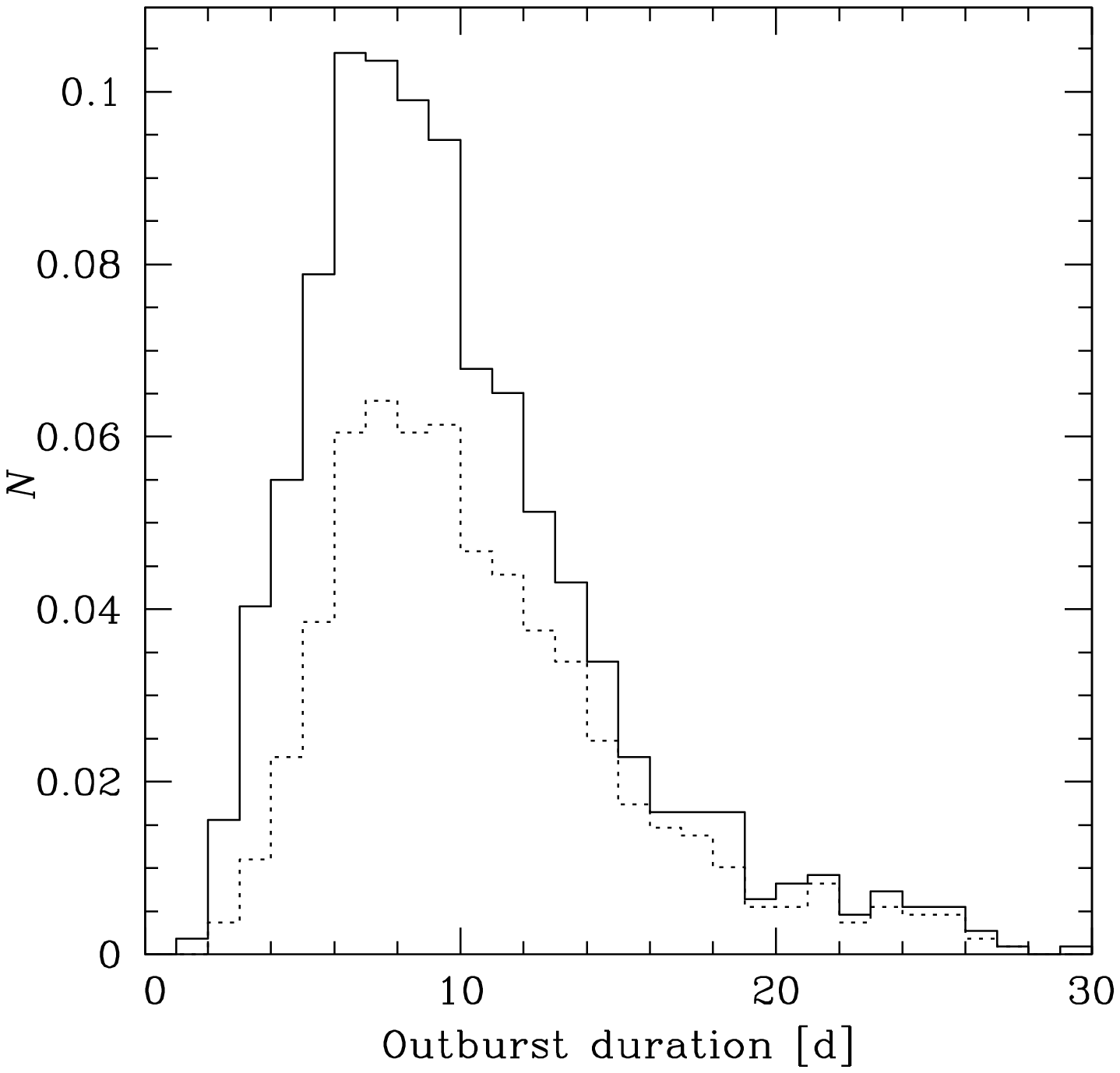}
\FigCap{{\it Left panel:} Distribution of the outburst frequency (for objects with at least two outbursts recorded). The peak at 0.7 yr$^{-1}$ is produced by DNe with two outbursts recorded in years 2010--2013 (OGLE-IV). {\it Right panel:} Distribution of the mean outburst duration. DNe discovered by the algorithm of Mr\'oz \etal (2013) are given by the dotted line.}
\end{figure}

\subsection{Outburst Statistics}

Light curves of all of the objects were automatically analyzed in order to investigate properties of individual outbursts. We assumed that the outburst starts/ends when the star is $3\sigma$ brighter than the mean brightness in quiescence ($\sigma$ is the dispersion of measurements in quiescence). If only one data point per outburst was detected, it had to be $5\sigma$ brighter than the mean. We visually inspected results of this automatic procedure for each star and found the very high success rate (>95\%). The procedure failed to identify properly outbursts for a few faint, low-amplitude objects (in that case we manually lowered the detection thresholds). For each star, we measured the number of observed outbursts, outburst frequency $\nu_{\rm obs}$ (corrected for seasonal gaps), and mean outburst duration $\langle D \rangle$. We show these numbers in Table~\tabOGLEDN{}, however, for faint low-amplitude objects they should be treated with caution.

We detected over a hundred outbursts for several U Gem-type DNe. Because objects have been observed for different time intervals, they should be compared using the outburst frequency $\nu_{\rm obs}$, corrected for seasonal gaps in the light curves (see left panel of Fig. \figCVstats{}). The outburst frequency can reach up to $\sim 20-25$ yr$^{-1}$ for some U~Gem-type DNe. On the other hand, only one outburst was recorded for over 200 DNe with duty cycles as short as 0.5\%.

The distribution of the mean outburst duration is shown in the right panel of Fig.~\figCVstats{}. In 90\% of cases $3.8 < \langle D \rangle < 22 $ d with a median value of 9.0 d. The majority of objects with outbursts longer than 20~d are WZ Sge-type DNe (Section~4.1).

\subsection{Orbital and Superhump Periods}

We analyzed quiescent light curves of all DN candidates from our sample to search for periodic variations (orbital humps, eclipses, etc). We searched periods in a range 0.003 -- 10 d using the Lomb-Scargle algorithm (Scargle 1982) and the analysis of variance method (Schwarzenberg-Czerny 1996). Orbital periods were measured for 26 objects. We also measured superhump periods for another several SU UMa-type DNe. A list of orbital and superhump periods is available online.

\subsection{X-ray Counterparts}

CVs, particularly systems with magnetic white dwarfs, may be strong sources of an X-ray emission. We cross-matched our objects with catalogs of X-ray sources from {\it Chandra} and {\it XMM-Newton} satellites, which are characterized by high angular resolution. For a search radius of $10''$, X-ray counterparts were found for 36 DNe. In 85 per cent of cases the offset is smaller than $3\farcs0$. X-ray counterparts were found in the {\it Chandra} Galactic bulge survey catalog (Jonker \etal 2011, 2014), {\it XMM-Newton} serendipitous source catalog (Rosen \etal 2015), and {\it XMM-Newton} slew survey catalog (Saxton \etal 2008). A full list is available online.
DNe with X-ray counterparts are on average 1 mag brighter compared to other systems, which suggests they are relatively close systems.

\section{Interesting Objects}

\subsection{WZ Sge-type Dwarf Novae}

WZ Sge-type dwarf novae form a subclass of SU UMa-type objects with rare (once per several years) large-amplitude superoutbursts. They usually have small mass ratios ($q \lesssim 0.1$) indicating that the secondary is a low-mass red dwarf or even a brown dwarf. Hence, some members of this class are believed to be period bouncers (Kato \etal 2013; Nakata \etal 2014), which are evolved CVs past the period minimum. According to Kato (2015), one of the best distinguishing characteristics of WZ Sge-type dwarf novae is the presence of early superhumps during early stages of superoutbursts (although their amplitude depends on the system geometry and they were not observed in some systems). In many cases the cadence of our observations prevents us from detecting this phenomenon. And thus, we classify as WZ Sge-type candidates objects fulfilling at least three of the following criteria: 1) rare (often only one), long (at least $\approx 20$ d), trapezoidal-shape superoutbursts, 2) lack of normal outbursts (except rebrightenings), 3) the presence of rebrightenings (``echo'' outbursts), and 4) large outburst amplitude. 

In Table \tabWZSge{}, we summarize basic properties of each system: maximal brightness $I_{\rm max}$, amplitude $\Delta I$, rebrightening type (following Imada \etal (2006) classification system), number of rebrightenings $N_{\rm reb}$, number of normal outbursts $N_{\rm n}$,  duration of superoutbursts $\Delta T$, superhump period $P_{\rm sh}$ (if measured). Example light curves are shown in Fig. \figWZSge{}.

Two systems with multiple ``echo'' outbursts have long superhump periods: OGLE-BLG-DN-0174 (0.14474(4)~d) and OGLE-BLG-DN-0595 (0.0972(1)~d), longer than in other similar objects. They might host an unusual, evolved secondary (Thorstensen \etal 2002; Nakata \etal 2013).

\begin{figure}[htb]
\includegraphics[width=0.49\textwidth]{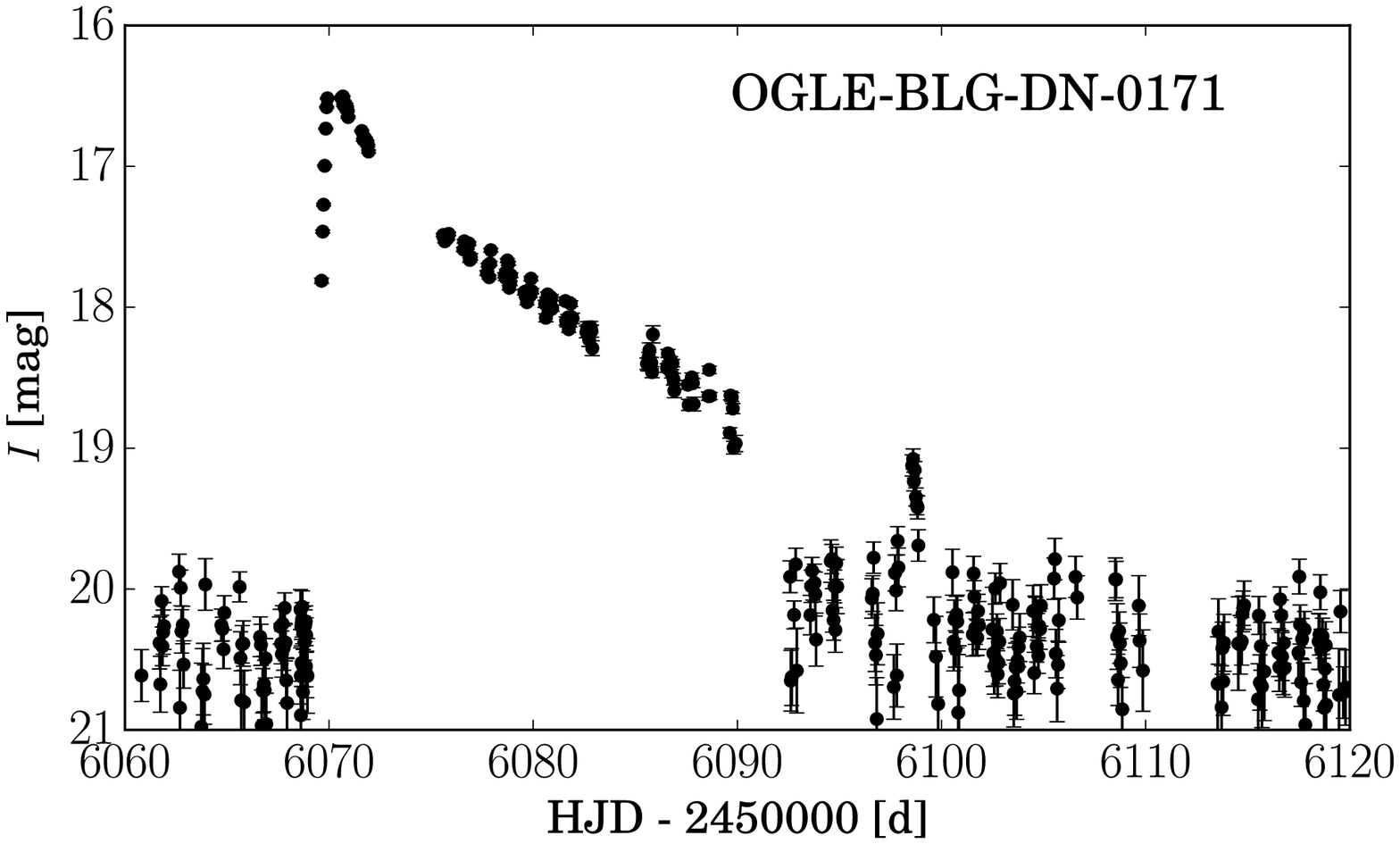} 
\includegraphics[width=0.49\textwidth]{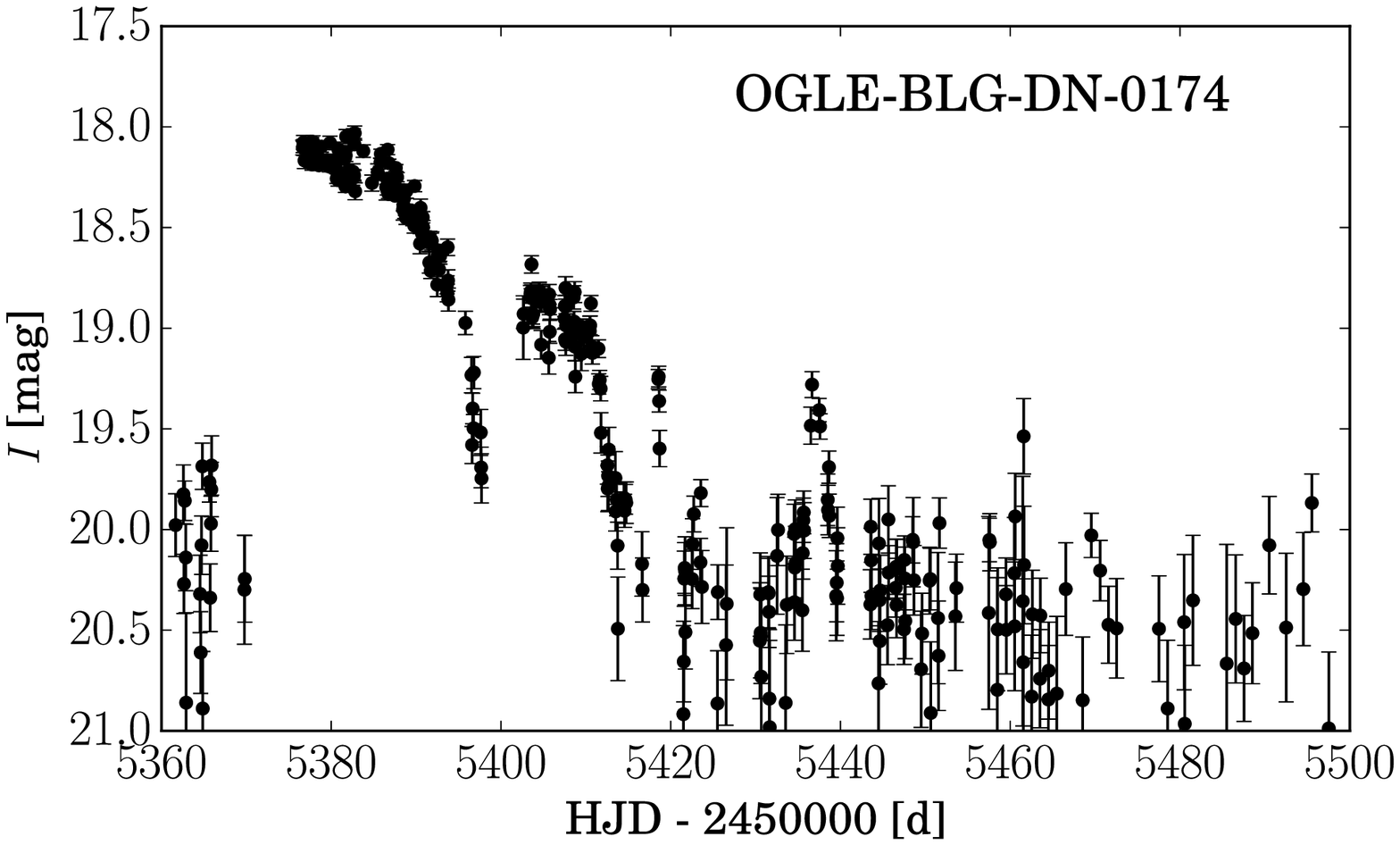} \\
\includegraphics[width=0.49\textwidth]{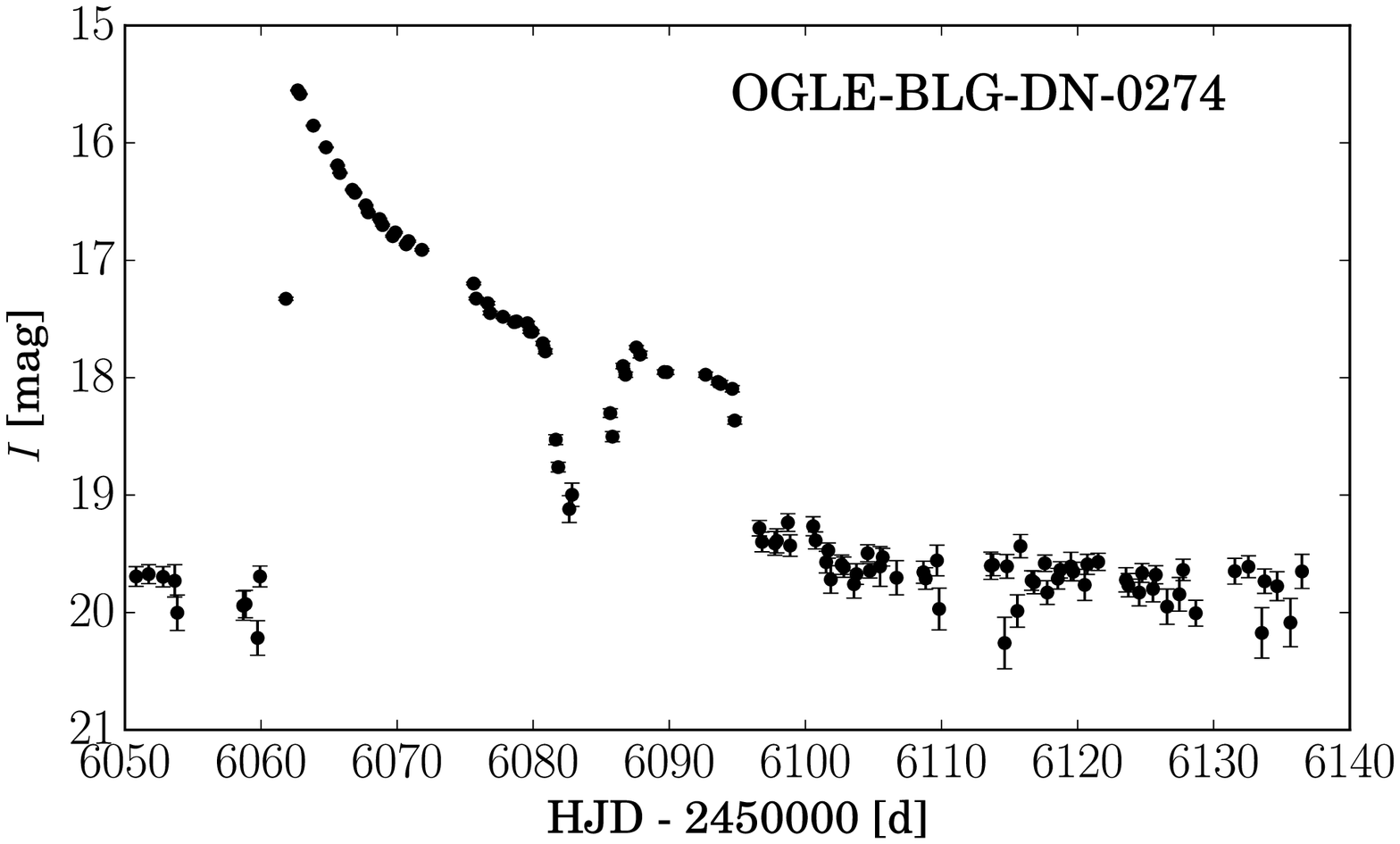}
\includegraphics[width=0.49\textwidth]{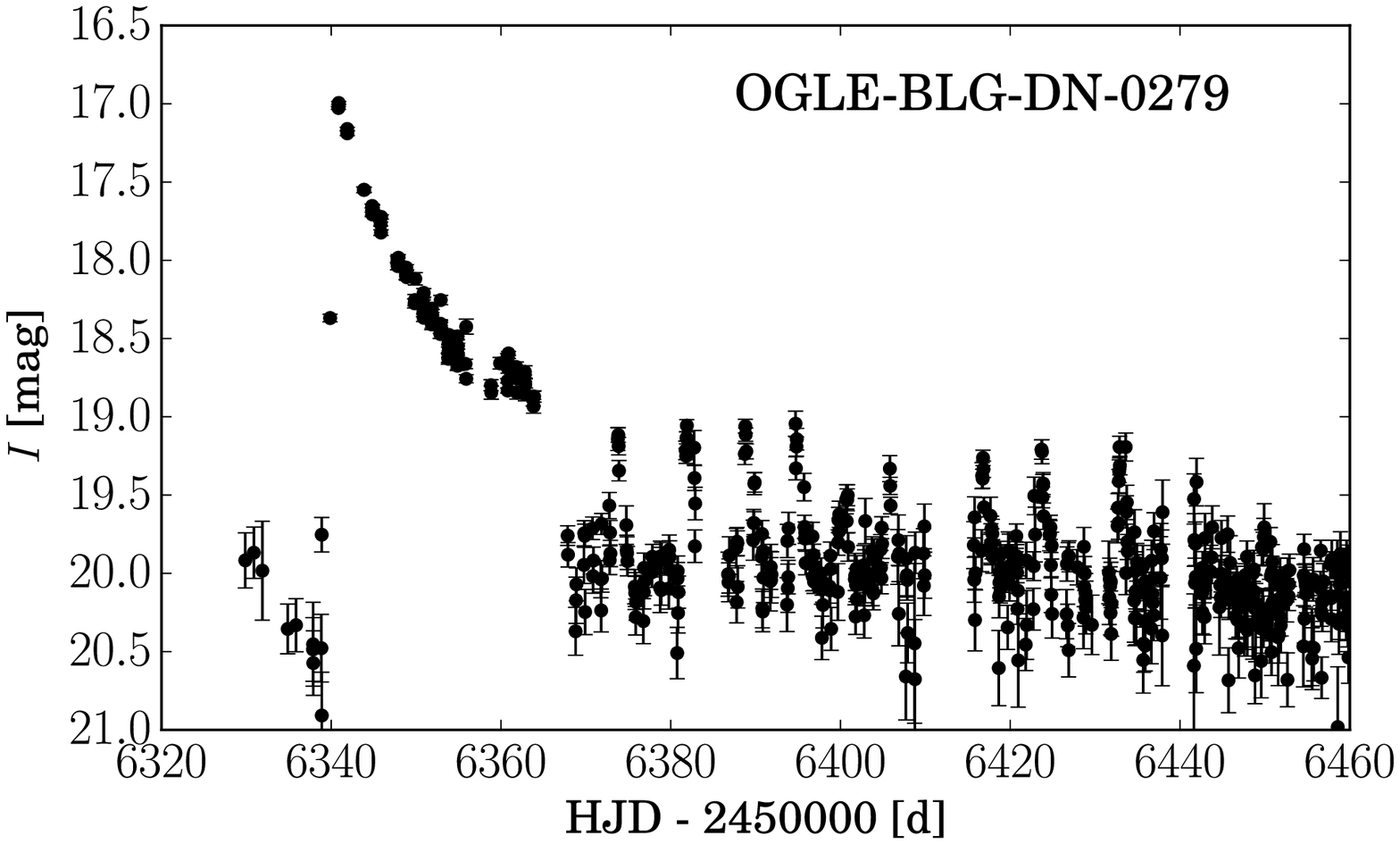} \\
\includegraphics[width=0.49\textwidth]{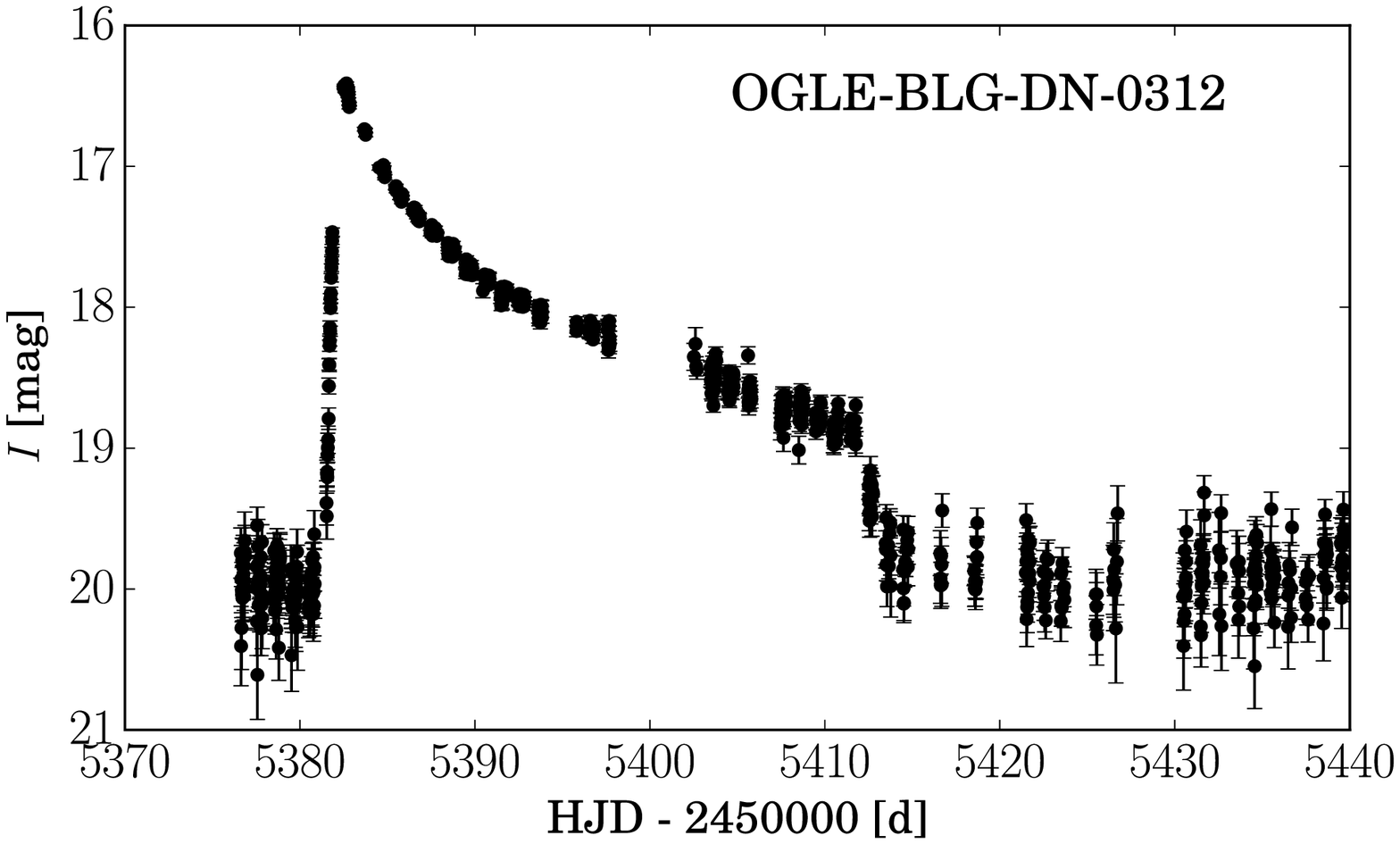}
\includegraphics[width=0.49\textwidth]{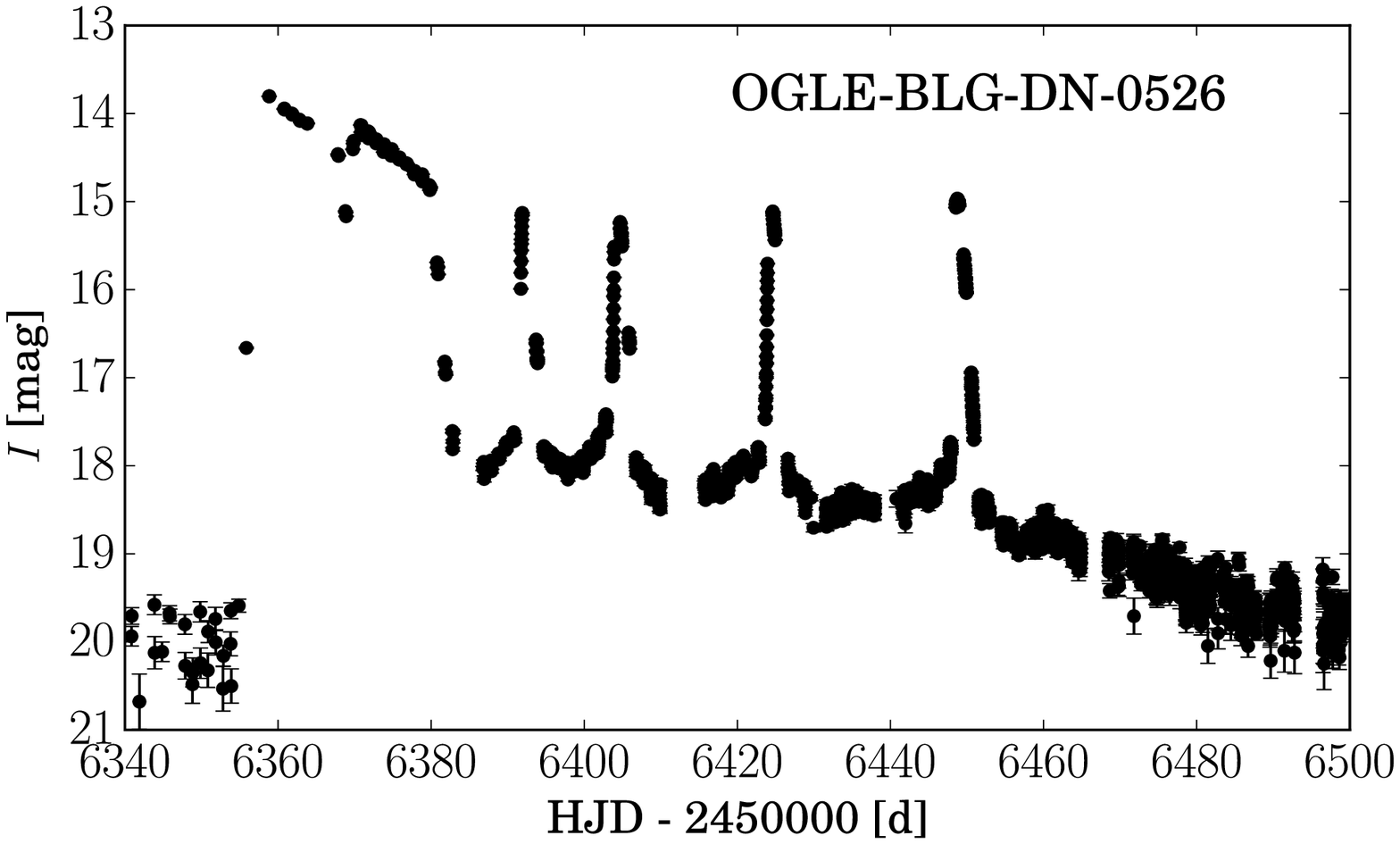} 
\FigCap{Examples of light curves of WZ Sge-type dwarf novae which show different types of outbursts: A - with long duration rebrightening (-0274), B - with multiple rebrightenings (-0174, -0279, -0526), C - with single short rebrightening (-0171), D - no rebrightenings (-0312).}
\label{fig:wz_sge}
\end{figure}

\setlength\tabcolsep{4pt}
\renewcommand{\TableFont}{\scriptsize}
\MakeTable{@{} lr@{\upd}r@{\upm}r@{\fs}lr@{\arcd}r@{\arcm}r@{\farcs}lccrrrrl @{}}{\textwidth}{Photometric data on the WZ Sge type candidates (first 15 objects).}
{
\hline
\multicolumn{1}{c}{Name} & \multicolumn{4}{c}{RA} & \multicolumn{4}{c}{Decl.} & $I_{\rm max}$ & $\Delta I$ & Type & $N_{\rm reb}$ & $N_{\rm n}$ & \multicolumn{1}{c}{$\Delta T$} & \multicolumn{1}{c}{$P_{\rm sh}$} \\
& \multicolumn{4}{c}{J2000.0} & \multicolumn{4}{c}{J2000.0} & [mag] & [mag] & & & & \multicolumn{1}{c}{[d]} & \multicolumn{1}{c}{[d]} \\
\hline
OGLE-BLG-DN-0009 & 17 & 16 & 02 & 98 & -29 & 10 & 10 & 1 & 16.47 & 1.67 & A & 1 & 0 & 11 \\
OGLE-BLG-DN-0050 & 17 & 37 & 18 & 60 & -29 & 41 & 32 & 3 & 15.11 & 4.80 & B & 5 & 0 & 27 \\
OGLE-BLG-DN-0055 & 17 & 38 & 28 & 01 & -34 & 53 & 20 & 7 & 17.92 & 2.17 & - & - & 0 & 25 \\
OGLE-BLG-DN-0087 & 17 & 43 & 08 & 01 & -34 & 19 & 28 & 4 & 17.09 & 2.18 & D & 0 & 0 & 20 \\
OGLE-BLG-DN-0088 & 17 & 43 & 08 & 34 & -27 & 44 & 38 & 8 & 17.08 & 3.24 & C & 1 & 0 & 19 \\
OGLE-BLG-DN-0109 & 17 & 45 & 31 & 91 & -22 & 21 & 45 & 1 & 16.69 & 3.68 & D & 0 & 0 & 14 \\
OGLE-BLG-DN-0114 & 17 & 46 & 07 & 03 & -33 & 30 & 42 & 8 & 16.56 & 3.22 & A & 1 & 0 & 22 \\
OGLE-BLG-DN-0115 & 17 & 46 & 15 & 97 & -35 & 12 & 04 & 3 & 17.29 & 1.12 & AB & - & 0 & 24 \\
OGLE-BLG-DN-0132 & 17 & 47 & 46 & 47 & -23 & 56 & 44 & 8 & 16.90 & 3.51 & AB & - & 0 & 26 \\
OGLE-BLG-DN-0145 & 17 & 49 & 15 & 62 & -22 & 45 & 34 & 9 & 16.33 & 3.81 & A & 1 & 0 & 24 \\
OGLE-BLG-DN-0156 & 17 & 49 & 49 & 34 & -21 & 22 & 11 & 4 & 16.63 & 3.80 & D & 0 & 0 & 27 \\
OGLE-BLG-DN-0168 & 17 & 50 & 24 & 49 & -29 & 51 & 27 & 1 & 17.98 & 2.18 & AB & - & 0 & 24 \\
OGLE-BLG-DN-0171 & 17 & 50 & 28 & 71 & -30 & 46 & 36 & 8 & 16.50 & 3.94 & C & 1 & 0 & 22 \\
OGLE-BLG-DN-0174 & 17 & 50 & 43 & 71 & -28 & 35 & 00 & 5 & 18.03 & 2.36 & B & 3 & 1 & 22 & 0.14474(4)\\
OGLE-BLG-DN-0181 & 17 & 51 & 01 & 17 & -29 & 14 & 39 & 7 & 18.14 & 2.16 & C & 1 & 0 & 23 \\
\hline
\multicolumn{12}{p{9cm}}{The full table is available online.}
}
\setlength\tabcolsep{5pt}

\subsection{SU UMa-type Stars with ``Early'' Precursors}

For several SU UMa-type objects, we observed ``early'' precursor outbursts, preceding superoutbursts by 5--10 days (Fig. \figSUUMa{}). In some objects (like OGLE-BLG-DN-0185) they were present before each superoutburst. Osaki and Kato (2014) noticed similar features in the {\it Kepler} light curves of V1504~Cyg and V344~Lyr. They observed that superhumps appeared during the descending branch of the precursor outburst, which supports the thermal-tidal instability model of accretion disks in SU UMa-type DNe. Light curves of SU UMa-type stars with ``early'' precursors from our sample can be used to study the development of superhumps and to test the thermal-tidal instability model (Osaki and Kato 2013, 2014, Smak 2013, and references therein).

\begin{figure}[htb]
\includegraphics[width=0.49\textwidth]{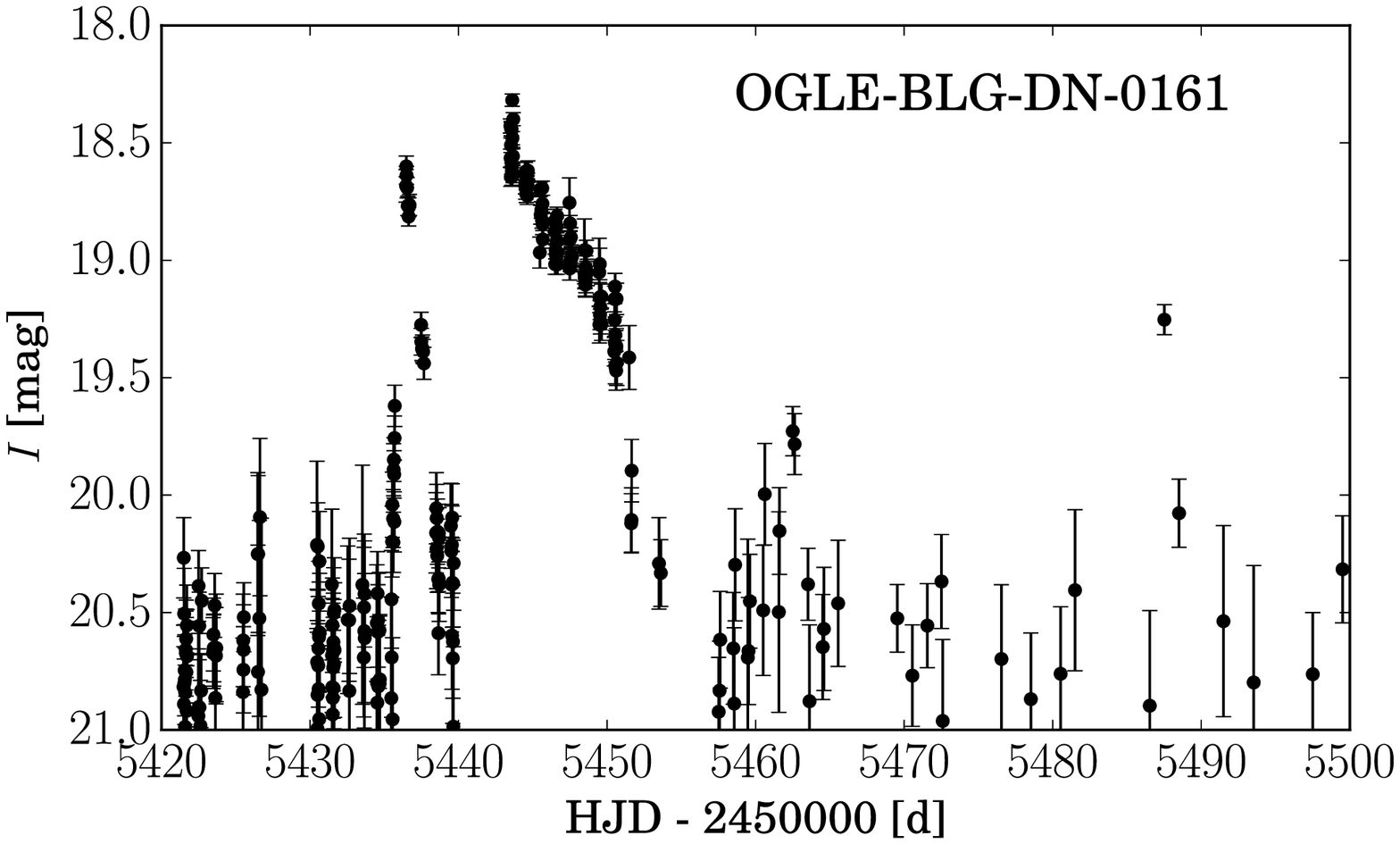}
\includegraphics[width=0.49\textwidth]{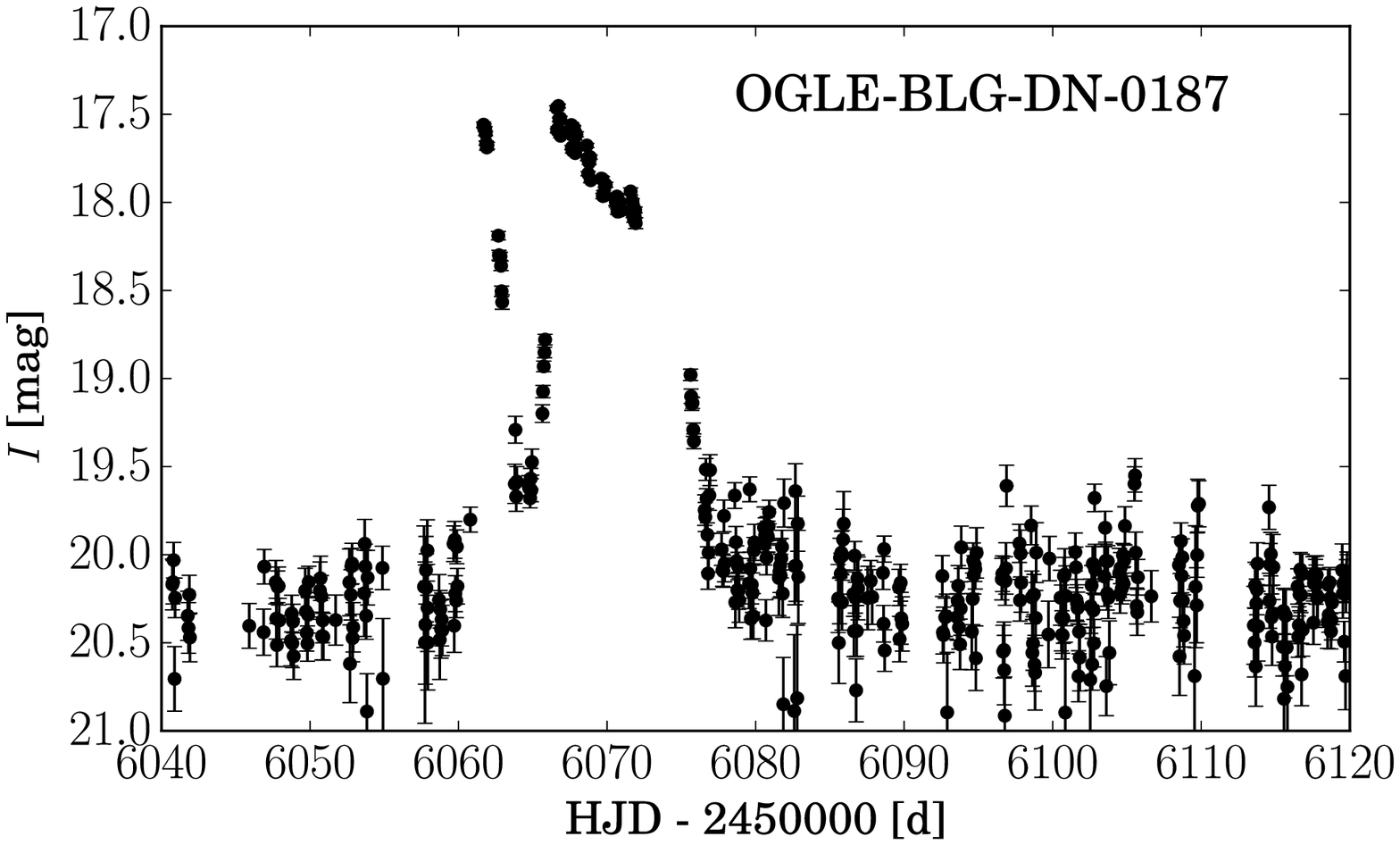}
\FigCap{Examples of light curves of SU UMa-type dwarf novae with ``early'' precursor outbursts.}
\end{figure}

\subsection{Z Camelopardalis-type Stars}

Z Camelopardalis-type dwarf novae\footnote{sites.google.com/site/thezcamlist/the-list} are characterized by standstills, during which outbursts cease and the brightness remains practically constant. This is attributed to variations of the mass-transfer rate in these systems: during standstills mass-transfer rate exceeds the critical value and the accretion disk is in stable configuration (no outbursts). If the mass-transfer rate drops, instabilities in the disk cause quasi-periodic outbursts (\eg Warner 1995). 

Five Z Cam-type objects were identified in our sample (Tab. \tabZCam{}). Among them, OGLE-BLG-DN-0219 (Fig. \figZCam{}) seems to be particularly interesting, it showed low-amplitude outbursts during standstills. Such behavior is very rare, it was observed only in a few other systems (Simonsen 2011; Szkody \etal 2013). According to Hameury \& Lasota (2014) it might be caused by mass transfer outbursts, although their origin remains unknown. The quiescent brightness varies (by $\Delta V = 1.2$ mag), suggesting significant mass-transfer rate variations in the unstable state.

\begin{figure}[htb]
\includegraphics[width=\textwidth]{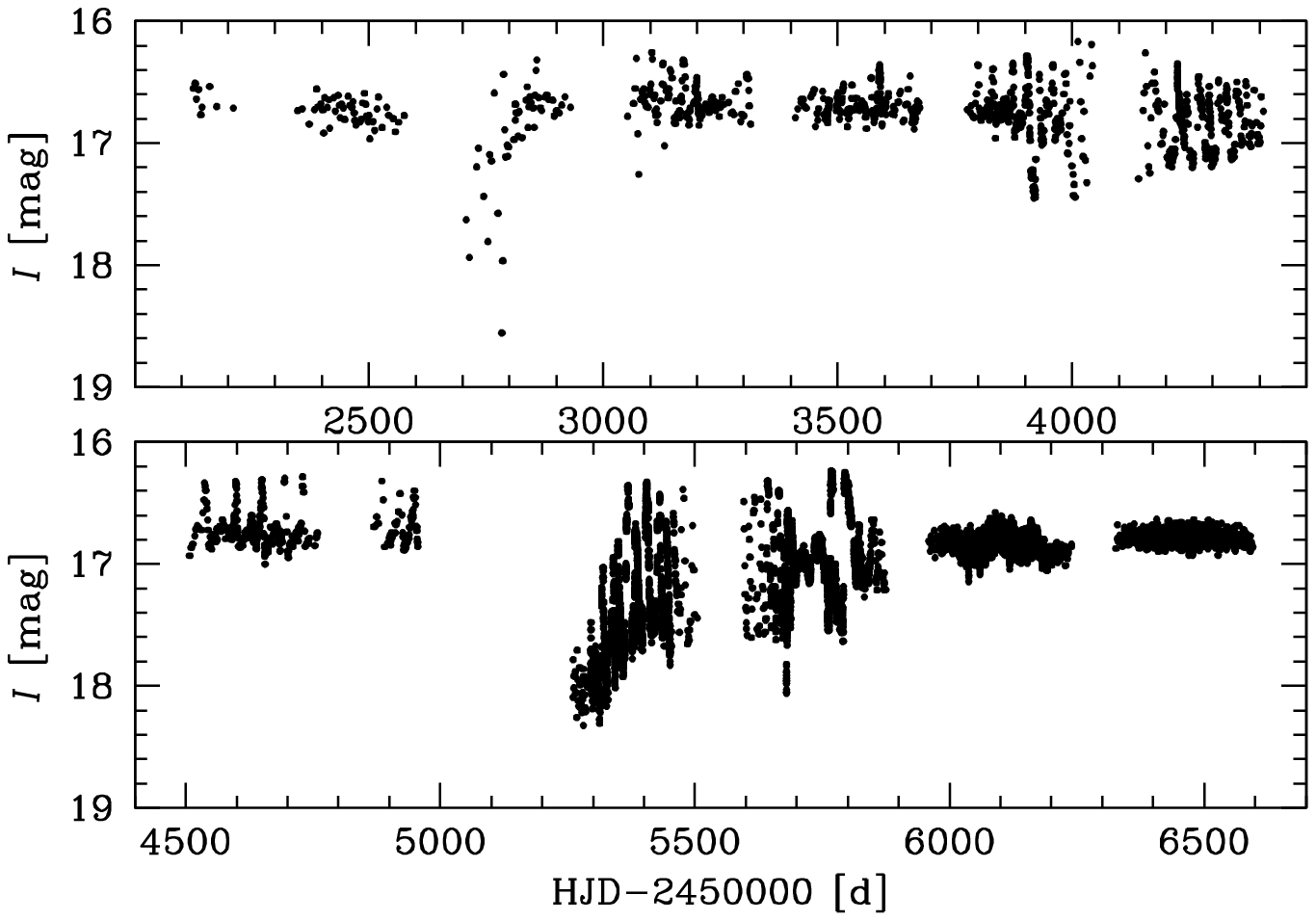}
\FigCap{Light curve of Z Cam-type star OGLE-BLG-DN-0219, showing outbursts during standstills. The quiescence brightness varies with time, suggesting significant variations of the mass-transfer rate.}
\end{figure}

\renewcommand{\TableFont}{\scriptsize}
\MakeTable{lr@{\uph}r@{\upm}r@{\fs}lr@{\arcd}r@{$'$}r@{\arcm}lccrrr}{\textwidth}{Photometric data on the Z Cam type stars.}
{
\hline
\multicolumn{1}{c}{Name} & \multicolumn{4}{c}{RA} & \multicolumn{4}{c}{Decl.} & $I_{\rm min}$ & $I_{\rm max}$ & $I_{\rm stand}$ & $T_{\rm out}$ & $D_{\rm out}$ \\
 & \multicolumn{4}{c}{J2000.0} & \multicolumn{4}{c}{J2000.0} & [mag] & [mag] & [mag] & [d] & [d] \\
\hline
OGLE-MC-DN-0017  & 04 & 46 & 53 & 58 & -67 & 23 & 57 & 5 & 18.76       & 16.90 & 17.70 & 15 & 10\\
OGLE-BLG-DN-0219 & 17 & 52 & 36 & 05 & -29 & 19 & 39 & 9 & 17.10-18.03 & 16.17 & 16.80 & 20 & 8-12 \\
OGLE-BLG-DN-0304 & 17 & 55 & 57 & 11 & -29 & 13 & 35 & 6 & 19.20       & 17.90 & 18.60 & 15 & 8-10 \\
OGLE-BLG-DN-0326 & 17 & 56 & 52 & 89 & -30 & 02 & 07 & 1 & 20.15       & 18.80 & 19.68 & 22 & 5-10 \\
OGLE-BLG-DN-0502 & 18 & 02 & 06 & 60 & -32 & 19 & 04 & 1 & 21.35       & 18.66 & 20.44 & 53 & 5-10 \\
\hline
}

\begin{figure}[htb]
\includegraphics[width=\textwidth]{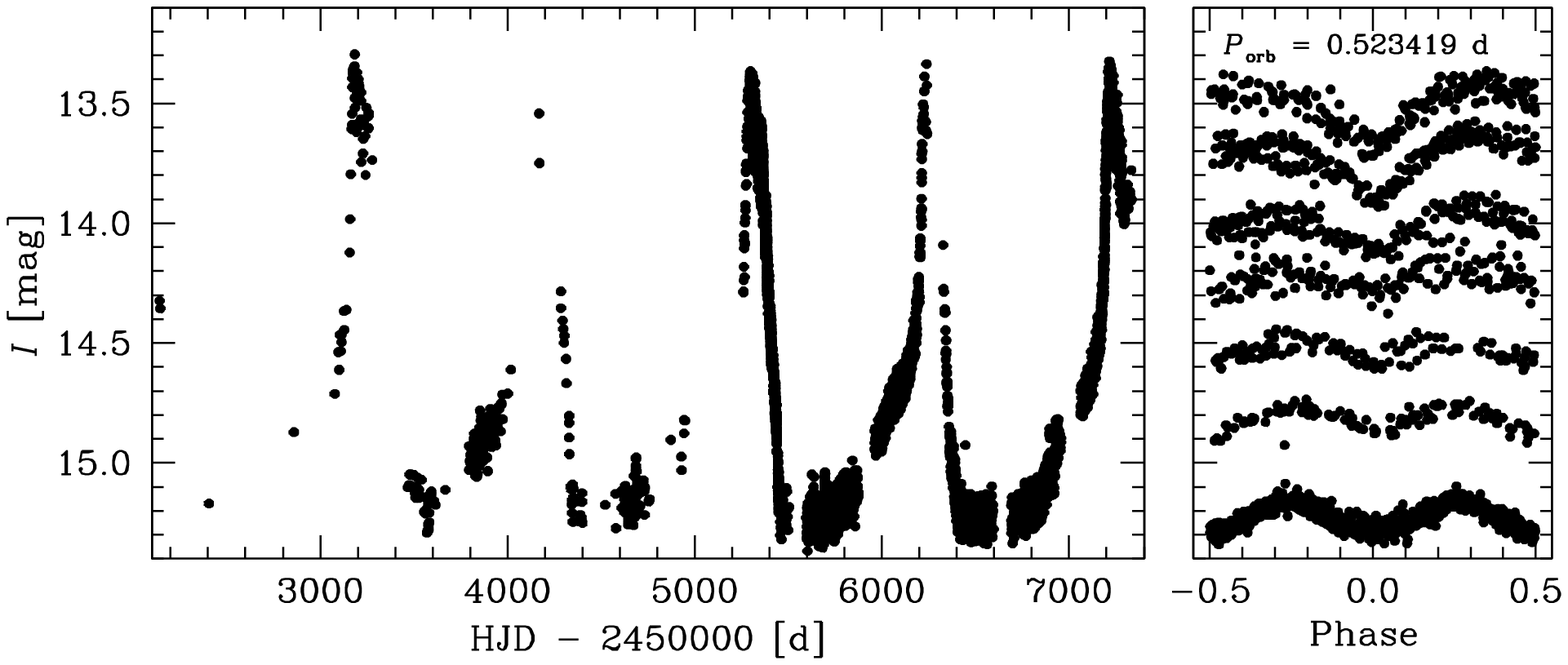}
\FigCap{Light curve of OGLE-BLG504.12.201843. Right panel shows light curve chunks phased with the orbital period $P_{\rm orb} = 0.523419(1)$ d -- their shape evolves as the outburst progresses.}
\end{figure}

\subsection{OGLE-BLG504.12.201843}

Although the light curve of OGLE-BLG504.12.201843 (Fig. \figBLG{}) does not clearly resemble that of DN, we dedicate a section to this object, because it seems to be a compact binary system as inferred from the variability in quiescence. High amplitude (almost two mag in $I$) outbursts recur on a timescale of 950--1020 d, but they are not strictly periodic. The outbursts shape is quite peculiar: in many transients, the rise to the maximum is very fast, while the decline to quiescence is several times longer. In OGLE-BLG504.12.201843, we observe slow, gradual brightening (lasting $\sim 1$ year), followed by quick ($\sim$ several days) rise to maximum, and then quite fast ($\sim 150$ d) decline to quiescence. Subsequent outbursts are very similar, but not identical.

In quiescence, the source lies blueward of the main-sequence on the color-magnitude diagram ($V-I=1.10$, $I=15.20$ mag), which may indicate the presence of an accretion disk. During outbursts the source becomes slightly bluer ($V-I=0.68$ mag). 
The interstellar reddening in the object direction (RA = 17\uph57\upm19\fs65, Decl. = -28\arcd08\arcm15\farcs7 for the epoch J2000.0) is moderate: $E(V-I) = 1.56 \pm 0.16$ (Nataf \etal 2013), $E(J-K_s) = 0.52 \pm 0.12$ (Gonzalez \etal 2012) as measured from the centroid of the red giant clump. However, the object has to be much closer than red clump stars (Galactic bulge), because the de-reddened color $(V-I)_0 \approx -0.5$ is unphysical. 

Both in quiescence and during outbursts, we detected photometric variability with the period of 0.523419(1) d, which we interpret as the orbital period. The shape of the light curve evolves as the outburst progresses (Fig. \figBLG{}). It resembles a small-amplitude sinusoid in quiescence (which might be caused by ellipsoidal variability) and then minima (``eclipses'') change both shape and depth. 

We speculate that this system might be a triple: a contact binary with the orbital period of 0.523419 d and a third body on a larger ($\sim 2.7$ yr) orbit. However, an $O-C$ analysis implies that the orbital period does not change at a level of $1-2\%$. 

\newpage
\section{K2 Campaign 9}

A field of view of $\approx 3.4$ deg$^2$ toward the Galactic bulge will be extensively monitored during the repurposed {\it Kepler} satellite mission (K2, Howell \etal 2014, Henderson \etal 2015) Campaign 9. The main goal of the Campaign 9 is the gravitational microlensing experiment, which with simultaneous ground-based observations should allow the measurement of microlensing parallaxes, and hence distances and masses of lenses. This campaign will differ from other K2 campaigns, because of specifications of the microlensing experiment. All pixels from a $\approx 3.4$ deg$^2$ region (``superstamp'', the shaded area in Fig. \figCVdistr{}) will be downloaded. The campaign will start in April 2016 and will last for 85 days. The cadence of {\it Kepler} observations will be 30 min. The K2C9 superstamp will also be monitored by a number of ground-based facilities, including the OGLE-IV Survey. Details can be found in Henderson \etal (2015).

130 DNe from our sample are located within the K2C9 superstamp (version of December 31st, 2015). First, they might contaminate microlensing detections, as outbursts of frequently erupting DNe can sometimes mimic microlensing events. Our list will allow rejecting obvious DNe at an early stage of an outburst.
Secondly, studies of individual targets located within the K2C9 superstamp will be possible (as in the case of V344~Lyr and V1504~Cyg, which were observed during the main {\it Kepler} mission, \eg Osaki \& Kato 2013). The complete list of CVs which will be observed during the K2 Campaign 9 is available online.

\section{CVOM: OGLE-IV Real Time Monitoring of Cataclysmic Variables}

We present the new OGLE-IV real-time data analysis system: CVOM, which has been designed to provide continuous real time photometric monitoring of selected CVs. The system is based on the XROM and RCOM, which have been used to monitor in the real-time optical counterparts of X-ray sources and R CrB-type variables, respectively (Udalski 2008). 

CV outbursts are unpredictable and the OGLE CVOM system can be used to trigger full variety of follow-up observations. The initial sample contains forty objects: the brightest SU UMa-type DNe from this paper and several old novae. The list of monitored objects can be easily extended.

The interactive access to the CVOM objects is provided via:
\begin{center}
{\it
http://ogle.astrouw.edu.pl/ogle4/cvom/cvom.html
}
\end{center}

The photometry can be downloaded from the OGLE archive:
\begin{center}
{\it
ftp://ftp.astrouw.edu.pl/ogle/ogle4/cvom
}
\end{center}

\Acknow{We would like to thank Profs. M. Kubiak and G. Pietrzy{\'n}ski, former members of the OGLE team, for their contribution to the collection of the OGLE photometric data over the past years. We thank T. Mazewski for his initial work on DNe in OGLE data. P.M. is supported by the ``Diamond Grant'' No. DI2013\slash014743 funded by the Polish Ministry of Science and Higher Education. The OGLE project has received funding from the National Science Center, Poland, grant MAESTRO 2014/14/A/ST9/00121 to A.U. This work has been supported by the Polish Ministry of Science and Higher Education through the program ``Ideas Plus'' award No. IdP2012 000162.}

\end{document}